\documentclass[10pt, aps, prd, amsmath, amssymb, floats, floatfix, twocolumn, notitlepage, superscriptaddress, nofootinbib, showpacs, longbibliography]{revtex4-2}
\usepackage{orcidlink}
\usepackage[T1]{fontenc}
\usepackage[utf8]{inputenc}
\usepackage{lmodern}
\usepackage{microtype}
\usepackage{amsmath,amssymb,amsfonts,bm,mathtools}
\usepackage{mathrsfs}
\usepackage{booktabs}
\usepackage{array}
\usepackage{tabularx}
\usepackage{xcolor}
\usepackage{hyperref}
\usepackage{siunitx}
\hypersetup{
  colorlinks=true,
  linkcolor=blue!45!black,
  citecolor=blue!45!black,
  urlcolor=blue!45!black,
  pdftitle={Horizon mergers and initial test-charge acceleration in charged black-hole geometries with anisotropic sources}
}
\hypersetup{colorlinks=true,linkcolor=black,citecolor=black,urlcolor=blue}
\newif\ifhighlight
\highlightfalse
\newcommand{\rev}[1]{\ifhighlight{\ifhighlight\color{blue}\fi#1}\else #1\fi}
\newcommand{\wq}{\omega_q}

\newcommand{\Rdiag}{\mathcal R}
\newcommand{\lmin}{\lambda_{\min}}
\newcommand{\lstar}{\lambda_*}

\newcommand{\dd}{\mathrm d}
\newcommand{\order}{\mathcal O}
\newcommand{\Msun}{M_\odot}

\begin{document}

\title{Horizon mergers and initial test-charge acceleration in charged black-hole geometries with anisotropic sources}

\author{Saeed Noori Gashti\orcidlink{0000-0001-7844-2640}}
\email{sn.gashti@du.ac.ir}
\affiliation{School of Physics, Damghan University, P. O. Box 3671641167, Damghan, Iran}
\affiliation{Center for Theoretical Physics, Khazar University, 41 Mehseti Street, Baku, AZ1096, Azerbaijan}

\author{ Mohammad Ali S. Afshar\orcidlink{0009-0001-3133-5992}}
\email{m.a.s.afshar@gmail.com}
\affiliation{ Department of Physics, Faculty of Basic Sciences, University of Mazandaran P. O. Box 47416-95447, Babolsar, Iran}
\affiliation{School of Physics, Damghan University, P. O. Box 3671641167, Damghan, Iran}
\affiliation{Center for Theoretical Physics, Khazar University, 41 Mehseti Street, Baku, AZ1096, Azerbaijan}

\author{ Mohammad Reza Alipour\orcidlink{0000-0001-8074-7865}}
\email{mohamad.alipour.1994@gmail.com; mr.alipour@stu.umz.ac.ir}
\affiliation{School of Physics, Damghan University, P. O. Box 3671641167, Damghan, Iran}
\affiliation{Center for Theoretical Physics, Khazar University, 41 Mehseti Street, Baku, AZ1096, Azerbaijan}

\begin{abstract}
We study horizon mergers and the initial radial acceleration of charged test particles in static,
spherically symmetric black-hole geometries with a string cloud, a Kiselev power-law component,
or a logarithmic profile conventionally. After separating the Maxwell contribution,
the effective environmental stress tensors are anisotropic and are not identified with homogeneous
quintessence or pressureless cold dark matter. Using a common mass-function formulation, we
derive the horizon-merger conditions and classify the Kiselev branches. In this family, the
inner-event merger charge provides only an upper bound: three distinct positive horizons
can additionally require a positive lower charge bound associated with the event-cosmological
merger. For a test particle released from rest relative to a static observer, we derive the local charge-to-mass threshold for outward initial acceleration. Writing $\lambda=q/\mu$ and
$z=Q/Q_{\rm ext}$, where $Q_{\rm ext}$ is the inner-event merger charge, the subextremal
string-cloud threshold has infimum $1/z$, approached only at infinity. For the subextremal
logarithmic profile, the threshold exceeds unity throughout the finite exterior and attains a finite global minimum. For $\alpha/M=0.4$ and $Q/M=1.2$, this minimum
is $\lambda_*\simeq1.251125$.
In a nondegenerate Kiselev static patch, the neutral balance radius is unique and radially unstable. Density matching supplies conditional profile normalizations, not observational fits. For the logarithmic profile, the integration parameter is distinguished from the mass combination inferred from circular motion. These results characterize geometric horizon bounds and test-particle dynamics; they do not establish a two-black-hole force law or a mechanism for cosmic acceleration, nor do they constitute a test of the Weak Gravity Conjecture.
\end{abstract}

\maketitle

\section{Introduction}\label{sec:intro}

The possible role of compact objects in late-time cosmic evolution has motivated several distinct lines of inquiry. One proposal is that the masses of black holes evolve with the cosmological scale factor, often parametrized as $M\propto a^k$ [1]. A separate line of work studies charged black holes surrounded by prescribed matter distributions, including a cloud of strings, a Kiselev power-law source, or a logarithmic profile commonly called perfect-fluid dark matter (PFDM) [5-9]. In these geometries the parameter denoted by $M$ need not be an asymptotic mass, and a horizon may persist even when the nominal ratio $Q/M$ exceeds unity.

These questions are related only at a broad level and require distinct diagnostics. Horizon existence is determined by the roots of the metric function. The acceleration of a charged test body is a local dynamical question. The interaction between two black holes requires a two-centre solution or a controlled approximation to one. The Weak Gravity Conjecture (WGC), in turn, is a statement about a properly normalized gauge sector and its charged spectrum [2,3]. A metric ratio such as $Q/M$ does not by itself test the conjecture.

\rev{Recent work has extended these themes to several related, but distinct, settings. Charged-particle epicyclic motion and circular-orbit stability have been studied for weakly magnetized Kiselev geometries [25], while geodesic motion and tidal observables have been examined in generalized Letelier-Alencar string-cloud spacetimes [26]. Other recent contributions consider charged string-cloud black holes in bumblebee gravity [27], logarithmic PFDM terms in asymptotically AdS black strings [28], and scalar-wave propagation in Kiselev backgrounds [29]. These models differ in their symmetry, matter content, or gravitational sector and are not direct substitutes for the Einstein-Maxwell spherical metrics adopted here. The present work instead isolates the branch structure of the prescribed metrics and the initial areal-radius acceleration of charged test bodies; it does not claim to determine epicyclic frequencies, ISCO stability, tidal tensors, or wave spectra.}

\rev{Throughout this work, Kiselev component refers to the static power-law source associated with the Kiselev metric. Its parameter $\wq$ will be shown to characterize the ratio of average principal pressure to density, not an isotropic perfect-fluid equation of state. The abbreviation PFDM is retained as the conventional name of the logarithmic metric profile; it does not imply isotropic pressure or a pressureless cold-dark-matter halo [23,24].}

Our purpose is to identify which conclusions follow directly from the prescribed static geometries. We first write the metrics in terms of a common mass function and derive their horizon-merger conditions. We then retain the radial diagnostic $[Q/m(r)]^2$ only as an ancillary algebraic quantity and derive the charged-particle equation of motion from the conserved canonical energy. The analysis distinguishes simple horizons from degenerate merger limits. We also prove uniqueness and radial instability of the neutral force-balance radius in the nondegenerate Kiselev static patch, and determine the global numerical threshold for outward acceleration in a representative PFDM geometry.

\rev{The density comparisons are recast as conditional normalizations at chosen reference radii. For the logarithmic profile, we explicitly distinguish the integration parameter from the mass inferred from circular motion. The cosmological-coupling discussion is kept separate from the static test-particle calculation; comparisons between the Kiselev parameter and a cosmological CPL parameter are not treated as statistical tests. Separate symbolic and regression tests provide implementation-level consistency checks for selected identities and reference cases; they are not an independent audit of the full manuscript.}

The paper is organized as follows. Section~\ref{sec:geometry} introduces the metrics and effective stress tensors. Section~\ref{sec:extremality} derives the extremality bounds and the Kiselev horizon criterion. Section~\ref{sec:forces} discusses the radial diagnostic and charged and neutral test-particle dynamics. Section~\ref{sec:density} gives conditional density normalizations. Section~\ref{sec:cosmo} addresses charge estimates, isotropy, and the charged-dust toy model. Section~\ref{sec:coupling} summarizes cosmological mass-coupling results without identifying them with the static-source parameter. Section~\ref{sec:conclusions} summarizes the restricted conclusions. Appendices provide the stress-tensor derivation, horizon proofs, test-particle dynamics, numerical procedure, unit conversions, and further checks. We use $G=c=4\pi\epsilon_0=1$ except where SI units are displayed explicitly.
\section{Geometries and effective mass functions}\label{sec:geometry}

We consider the static, spherically symmetric line element
\begin{equation}
 \begin{aligned}
 ds^2&=f(r)\,dt^2-f(r)^{-1}\,dr^2-r^2d\Omega^2,\\
 f(r)&=1-\frac{2m(r)}{r}+\frac{Q^2}{r^2}.
 \end{aligned}
 \label{eq:metric}
\end{equation}
The Maxwell contribution is written explicitly as $Q^2/r^2$. The function $m(r)$ is therefore the remaining mass-function term; it is not, in general, the Misner-Sharp mass or an ADM mass.

For a string cloud supplemented by a Kiselev component, and for the logarithmic PFDM profile, respectively, we take
\begin{align}
 m_{\rm sq}(r)&=M+\frac b2r+\frac\alpha2r^{-3\wq},
 \label{eq:msq}\\
 m_{\rm P}(r)&=M+\frac\alpha2\ln\!\left(\frac r\alpha\right).
 \label{eq:mpfdm}
\end{align}
The string-cloud geometry follows by setting the power-law term to zero. For the Kiselev component we assume $M>0$, $\alpha>0$, $b\geq0$, and $-1<\wq<-1/3$. For PFDM we take $\alpha>0$. The logarithmic profile is used as a phenomenological static metric profile, not as a globally valid cosmological halo solution.

\rev{It is useful to make the sign convention for the logarithmic term explicit. Equation~(\ref{eq:mpfdm}) gives}
\begin{equation}
 \rev{f_{\rm P}(r)=1-\frac{2M}{r}+\frac{Q^2}{r^2}
 -\frac{\alpha}{r}\ln\!\left(\frac r\alpha\right),\qquad \alpha>0.}
 \label{eq:pfdm-sign}
\end{equation}
\rev{With the environmental-source convention below, this profile has positive density $\rho_{\rm P}=\alpha/(8\pi r^3)$. If another convention writes the logarithmic term as $+\kappa r^{-1}\ln(r/|\kappa|)$, the same geometry corresponds to $\kappa=-\alpha$. Changing this sign without translating the source convention would define a different metric.}

\subsection{Effective stress tensor of the surrounding components}

\rev{In a static region we define the environmental stress tensor in an orthonormal frame by}
\begin{equation}
 \rev{T^{\hat a}{}_{\hat b\,{\rm env}}
 =\operatorname{diag}(\rho,-p_r,-p_t,-p_t).}
 \label{eq:environment-stress}
\end{equation}
\rev{The Maxwell contribution associated with the explicit $Q^2/r^2$ term in Eq.~(\ref{eq:metric}) is subtracted before the environmental quantities are identified. The Einstein equations then yield}
\begin{align}
 \rev{\rho(r)}&\rev{=\frac{m'(r)}{4\pi r^2},}\label{eq:density}\\
 \rev{p_r(r)}&\rev{=-\rho(r),\qquad
 p_t(r)=-\frac{m''(r)}{8\pi r}.}\label{eq:environment-pressures}
\end{align}
\rev{Consequently, a density profile does not by itself imply an isotropic-fluid interpretation. For the individual components these expressions give}
\begin{align}
 \ifhighlight\color{blue}\fi
 \rho_s&=\frac{b}{8\pi r^2},\quad
 p_{r,s}=-\rho_s,\quad p_{t,s}=0,
 \label{eq:string-stress}\\
 \rho_q&=\frac{\alpha p}{8\pi}r^{p-3},\quad
 p_{r,q}=-\rho_q,\quad
 p_{t,q}=-\frac{p-1}{2}\rho_q,
 \label{eq:kiselev-stress}\\
 \rho_{\rm P}&=\frac{\alpha}{8\pi r^3},\quad
 p_{r,\rm P}=-\rho_{\rm P},\quad
 p_{t,\rm P}=\frac12\rho_{\rm P},
 \label{eq:pfdm-stress}
\end{align}
\normalcolor
where $p=-3\wq\in(1,3)$.

\rev{Defining the average principal pressure by $\bar p=(p_r+2p_t)/3$, one obtains}
\begin{equation}
 \rev{\frac{\bar p_q}{\rho_q}=-\frac p3=\wq,\qquad
 \frac{\bar p_s}{\rho_s}=-\frac13,\qquad
 \bar p_{\rm P}=0.}
 \label{eq:average-pressure}
\end{equation}
\rev{Thus $\wq$ labels the average pressure of the anisotropic Kiselev source; it is not the equation-of-state parameter of an isotropic perfect fluid. The anisotropy is}
\begin{equation}
 \rev{p_{t,q}-p_{r,q}=\frac32(1+\wq)\rho_q,}
 \label{eq:kiselev-anisotropy}
\end{equation}
\rev{which is nonzero for $-1<\wq<-1/3$. Only in the limiting case $\wq=-1$ does the source become isotropic, with $p_r=p_t=-\rho_q$ and $\rho_q=3\alpha/(8\pi)$. In that limit $\alpha=\Lambda/3$ reproduces a cosmological constant.}

\rev{For the logarithmic PFDM profile, $\bar p_{\rm P}=0$ does not mean that the source is pressureless: its principal pressures satisfy $p_r=-\rho_{\rm P}$ and $p_t=\rho_{\rm P}/2$. The conventional name PFDM is therefore used only as a label for this metric profile. No identification with a cold, isotropic dark-matter halo is made.}

\rev{The environmental conservation equation is}
\begin{equation}
 \rev{p_r'+\frac{f'}{2f}(\rho+p_r)
 +\frac{2}{r}(p_r-p_t)=0,}
 \label{eq:environment-conservation}
\end{equation}
\rev{and is satisfied identically by Eqs.~(\ref{eq:density})-(\ref{eq:environment-pressures}). Appendix~\ref{app:stress} gives the direct component calculation and explains the limitation of isotropy within the one-function metric ansatz.}

\section{Extremality bounds and normalized charge ratios}\label{sec:extremality}

Horizons satisfy $f(r_h)=0$, equivalently
\begin{equation}
 g(r_h)=Q^2,\qquad g(r)=2m(r)r-r^2.
 \label{eq:horizon-g}
\end{equation}
At a degenerate horizon $r=r_e$, both $g(r_e)=Q_{\rm ext}^2$ and $g'(r_e)=0$ hold. The second condition is
\begin{equation}
 (rm)'\big|_{r_e}=r_e,
 \label{eq:extremality-condition}
\end{equation}
which gives
\rev{\begin{equation}
 Q_{\rm ext}^2
 =r_e\left[m(r_e)-r_e m'(r_e)\right].
 \label{eq:qext-general}
\end{equation}}

For $m=M$, these relations reduce to the Reissner-Nordstr\"om result $r_e=Q_{\rm ext}=M$.

\subsection{String cloud}

For $m=M+br/2$, the extremal radius and charge are
\begin{equation}
 r_e=\frac{M}{1-b},\qquad
 Q_{\rm ext}=\frac{M}{\sqrt{1-b}},\qquad 0\leq b<1.
 \label{eq:string-extremal}
\end{equation}
The horizon radii for $Q<Q_{\rm ext}$ are
\begin{equation}
 r_\pm=\frac{M\pm\sqrt{M^2-(1-b)Q^2}}{1-b}.
 \label{eq:string-roots}
\end{equation}
The small-$b$ expansion is
\begin{equation}
 \frac{Q_{\rm ext}}M-1=\frac b2+\frac{3b^2}{8}+\order(b^3).
 \label{eq:string-expansion}
\end{equation}

\subsection{Logarithmic PFDM profile}

For the logarithmic profile in Eq.~(\ref{eq:mpfdm}),
assume $M>0$, $\alpha>0$, and $Q>0$. The horizon function is
\begin{equation}
 g(r)=2Mr+\alpha r\ln\!\left(\frac{r}{\alpha}\right)-r^2.
\end{equation}
At a double horizon, $g'(r_e)=0$ and
$Q_{\rm ext}^2=g(r_e)$ give
\begin{equation}
 \begin{aligned}
 r_e&=M+\frac{\alpha}{2}
 \left[\ln\!\left(\frac{r_e}{\alpha}\right)+1\right],\\
 Q_{\rm ext}^2&=r_e(r_e-\alpha).
 \end{aligned}
 \label{eq:pfdm-extremal}
\end{equation}

The existence of a positive-charge black-hole branch requires $0<\alpha<2M$. To establish this restriction,
set $x=r/\alpha$ and use $\ln x\leq x-1$:
\begin{equation}
 \frac{g(r)}{r}
 =2M-\alpha+\alpha(\ln x-x+1)
 \leq2M-\alpha.
\end{equation}
Hence, if $\alpha\geq2M$, then $g(r)\leq0$ for all $r>0$, and the horizon equation $g(r)=Q^2$ has no solution for $Q>0$.

Conversely, if $0<\alpha<2M$, then
\begin{equation}
 g(\alpha)=\alpha(2M-\alpha)>0.
\end{equation}
Since $g(r)\to0$ from below as $r\to0^+$ and $g(r)\to-\infty$ as $r\to\infty$,
the function has a positive maximum. Moreover,
\begin{equation}
 g'(\alpha)=2M-\alpha>0,\qquad
 g''(r)=\frac{\alpha}{r}-2<0
 \quad (r\geq\alpha).
\end{equation}
Thus the maximum relevant to the black-hole branch occurs at the unique stationary radius $r_e>\alpha$,which selects the solution of Eq.~(\ref{eq:pfdm-extremal}).

For fixed $M$ and $\alpha$ satisfying $0<\alpha<2M$, the additional charge condition is
\begin{equation}
 0<Q\leq Q_{\rm ext}.
\end{equation}
For $0<Q<Q_{\rm ext}$, there are two distinct positive horizons; at $Q=Q_{\rm ext}$, they
merge into a double horizon. For $Q>Q_{\rm ext}$, no horizon exists.

For $\alpha/M\ll1$, the leading extremality shift is
\begin{equation}
 \frac{Q_{\rm ext}}{M}-1\simeq
 \frac{\alpha}{2M}\ln\!\left(\frac{M}{\alpha}\right).
 \label{eq:pfdm-expansion}
\end{equation}

\subsection{Kiselev component with a string cloud}

\rev{
The horizon classification in this subsection assumes $M>0$, $\alpha>0$, and $Q>0$.
The neutral case is treated separately where needed. For the combined model, let $p=-3\wq\in(1,3)$ and $A=1-b$. Then}
\begin{equation}
 \ifhighlight\color{blue}\fi
 \begin{aligned}
 g(r)&=2Mr-Ar^2+\alpha r^{p+1},\\
 H(r)&\equiv\frac{g'(r)}2
 =M-Ar+\frac{\alpha(p+1)}2r^p.
 \end{aligned}
 \label{eq:kiselev-gH}
\end{equation}
\normalcolor
\rev{For $A\leq0$, $H(r)>0$ and there is one positive root of $g=Q^2$. For $A>0$, $H$ has a unique minimum. Two distinct positive stationary radii exist precisely when}
\begin{equation}
 \rev{0<\alpha<\alpha_{\rm crit},\qquad
 \alpha_{\rm crit}=\frac{2A}{p(p+1)}
 \left[\frac{A(p-1)}{pM}\right]^{p-1}.}
 \label{eq:acrit}
\end{equation}
\rev{Denote these stationary radii by $r_a<r_b$; $g$ has a local maximum at $r_a$ and a local minimum at $r_b$. Define}
\begin{equation}
 \rev{Q_C^2=g(r_a),\qquad G_N=g(r_b),\qquad
 z\equiv\frac{Q}{Q_C}.}
 \label{eq:kiselev-bounds}
\end{equation}
\rev{The three-simple-positive-root domain is}
\begin{equation}
 \rev{\max\{0,G_N\}<Q^2<Q_C^2,\qquad 0<\alpha<\alpha_{\rm crit}.}
 \label{eq:kiselev-horizon-criterion}
\end{equation}
\rev{The upper boundary $Q=Q_C$ is the inner-event merger. If $G_N>0$, the lower boundary $Q^2=G_N$ is the event-cosmological merger; when $G_N\leq0$, no positive lower charge boundary is present. Thus $z<1$ is only an upper-charge condition and is not sufficient for a black-hole static patch. At $\alpha=\alpha_{\rm crit}$, the stationary radii coalesce. A triple horizon occurs only when the charge simultaneously satisfies $Q^2=Q_u^2$, as specified in Appendix~\ref{app:kiselev}. This limiting configuration is not an open nondegenerate static region.}

For $p=2$ the stationary radii and critical amplitude are explicit:
\begin{equation}
 r_{a,b}=\frac{A\mp\sqrt{A^2-6\alpha M}}{3\alpha},
 \qquad \alpha_{\rm crit}=\frac{A^2}{6M}.
 \label{eq:p2-stationary}
\end{equation}
\rev{An explicit counterexample to the sufficiency of $z<1$ is $M=1$, $b=0.1$, $\alpha=0.12$, $p=2$, and $Q=0.5$. Here $\alpha_{\rm crit}=0.135$, $r_a=5/3$, $r_b=10/3$, $Q_C^2=25/18$, and $G_N=10/9$. Although $z=0.424264<1$, one has $Q^2<G_N$ and only one positive horizon root, $r\simeq0.132795$. The root count is checked independently from the cubic horizon polynomial in Appendix~\ref{app:numerics}.}

\begin{figure}[t]
 \includegraphics[width=\columnwidth]{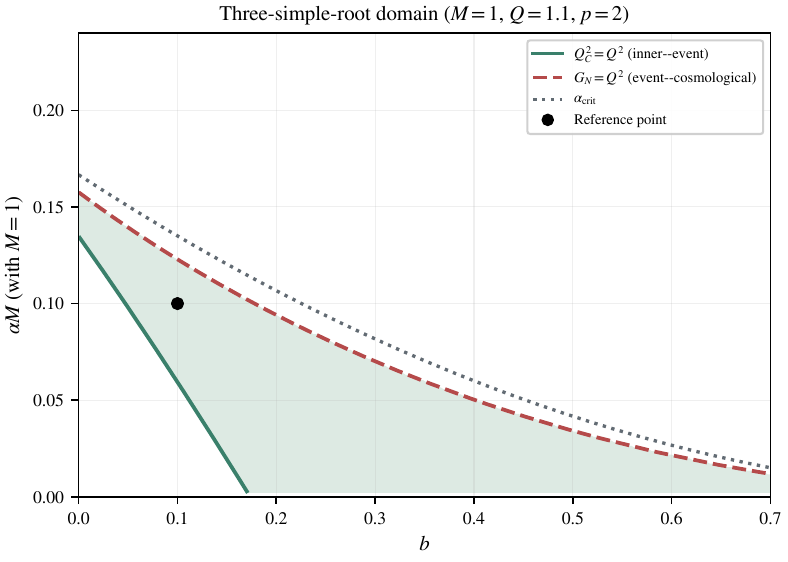}
 \caption{Horizon topology for $M=1$, $Q=1.1$, and $p=2$ in the $(b,\alpha M)$ plane. The shaded domain satisfies the three-simple-root criterion. The solid green and dashed red curves denote the inner-event and event-cosmological merger boundaries, respectively; the dotted curve is $\alpha_{\rm crit}$. The black point marks $(b,\alpha M)=(0.1,0.1)$.}
 \label{fig:kiselev-topology}
\end{figure}

Figure~\ref{fig:extremality} compares the exact merger shifts with their leading expansions. Figure~\ref{fig:kiselev-topology} displays the two-sided root criterion for the $p=2$, $M=1$, $Q=1.1$ slice. The shaded region is the domain with three distinct simple positive roots; merger curves are excluded from its interior.

\begin{figure*}[t]
 \includegraphics[width=0.95\textwidth]{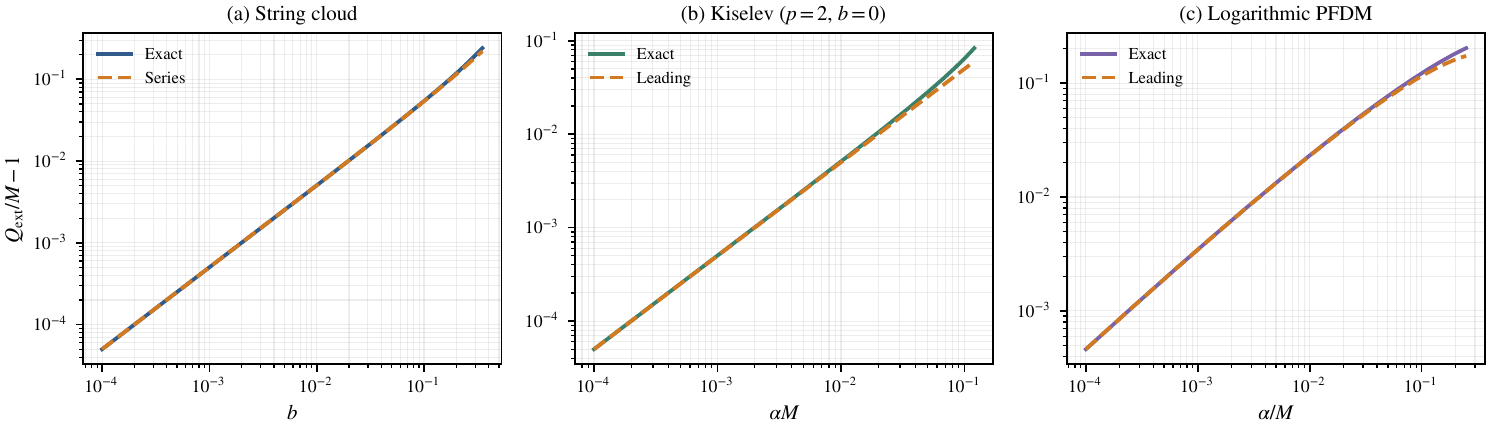}
 \caption{Extremality shifts for (a) a string cloud, (b) a Kiselev component with $p=2$, and (c) the logarithmic PFDM profile. Solid curves are obtained from the non-expanded extremality equations; dashed curves show the corresponding leading or second-order approximations. The horizontal and vertical quantities are dimensionless. The reference mass is $M=1$; panel (b) uses $b=0$ and the dimensionless amplitude $\alpha M$.}
 \label{fig:extremality}
\end{figure*}

\section{Radial diagnostics and test-particle dynamics}\label{sec:forces}

\subsection{An ancillary radial diagnostic}

\rev{For completeness, define the dimensionless quantity}
\begin{equation}
 \Rdiag(r)\equiv\left[\frac{Q}{m(r)}\right]^2.
 \label{eq:radial-proxy}
\end{equation}
\rev{This is an algebraic diagnostic constructed from the mass-function term after the Maxwell contribution has been separated. It is not a covariant force ratio. It should also be distinguished from the Misner-Sharp mass, which for Eq.~(\ref{eq:metric}) is}
\begin{equation}
 \rev{M_{\rm MS}(r)=\frac r2[1-f(r)]
 =m(r)-\frac{Q^2}{2r}.}
 \label{eq:ms-mass}
\end{equation}
\rev{A second combination,}
\begin{equation}
 N(r)\equiv m-rm'-\frac{Q^2}{r}=\frac{r^2f'}2,
 \label{eq:N-definition}
\end{equation}
\rev{controls the sign of the radial acceleration of a neutral body instantaneously at rest. It is neither $m(r)$ nor $M_{\rm MS}(r)$.}

At a horizon, Eq.~(\ref{eq:horizon-g}) implies
\begin{equation}
 Q^2=2m(r_h)r_h-r_h^2
 =m(r_h)^2-[m(r_h)-r_h]^2\leq m(r_h)^2.
 \label{eq:proxy-horizon-identity}
\end{equation}
On the horizon-bearing branches considered here $m(r_+)>0$ and $m'(r)\geq0$. Therefore $m(r)\geq m(r_+)\geq Q$ in the exterior, and
\begin{equation}
 \Rdiag(r)\leq1.
 \label{eq:proxy-bound}
\end{equation}
\rev{The statement is understood on the exterior closure: $r\geq r_+$ when no cosmological horizon is present, and $r_+\leq r\leq r_c$ when the static patch is bounded by a cosmological horizon. A static observer itself exists only in the open domain $r_+<r<r_c$. Equality throughout the exterior occurs in the extremal Reissner-Nordstr\"om limit; for the nontrivial environmental profiles considered here the inequality is strict away from the horizon boundary. Appendix~\ref{app:diagnostic} supplies the sign argument. Because no separate two-body observable is attached to Eq.~(\ref{eq:radial-proxy}), we use it only as a subsidiary geometric lemma.}

\begin{figure}[t]
 \includegraphics[width=\columnwidth]{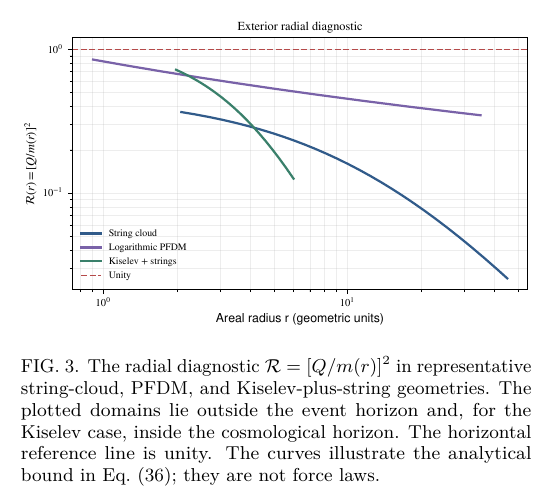}
 \caption{The radial diagnostic $\Rdiag=[Q/m(r)]^2$ in representative string-cloud, PFDM, and Kiselev-plus-string geometries. The plotted domains lie outside the event horizon and, for the Kiselev case, inside the cosmological horizon. The horizontal reference line is unity. The curves illustrate the analytical bound in Eq.~(\ref{eq:proxy-bound}); they are not force laws.}
 \label{fig:diagnostic}
\end{figure}

\subsection{Charged radial motion and the initial-acceleration threshold}

\rev{We now define the operational force question using the radial motion of a freely moving test body. Take the Maxwell potential in the gauge}
\begin{equation}
 \rev{A=A_t(r)\,dt,\qquad A_t(r)=\frac Qr,}
 \label{eq:gauge-potential}
\end{equation}
\rev{and let $\lambda=q/\mu$ for a particle of rest mass $\mu$ and charge $q$, with $Q>0$. An overdot denotes differentiation with respect to the particle's proper time $\tau$. The conserved canonical energy per unit rest mass is}
\begin{equation}
 \rev{\varepsilon=f\dot t+\lambda\frac Qr.}
 \label{eq:canonical-energy}
\end{equation}
\rev{The normalization of the four-velocity gives}
\begin{equation}
 \rev{\dot r^{\,2}=\left(\varepsilon-\lambda\frac Qr\right)^2-f(r).}
 \label{eq:radial-geodesic}
\end{equation}
\rev{At an event where the particle is instantaneously at rest relative to a static observer, $\dot r=0$ and $\varepsilon-\lambda Q/r=\sqrt f$ for a future-directed trajectory. Using the radial equation of motion, or its continuous limit at a turning point, gives the initial areal-radius acceleration $d^2r/d\tau^2$}
\begin{equation}
 \rev{\ddot r=-\frac{f'}2+\lambda\frac{Q\sqrt f}{r^2}.}
 \label{eq:radial-acceleration}
\end{equation}
\rev{Thus the initial acceleration is outward precisely when}
\begin{equation}
 \rev{\ddot r>0\quad\Longleftrightarrow\quad
 \lambda>\lmin(r)\equiv\frac{r^2f'}{2Q\sqrt f}
 =\frac{N(r)}{Q\sqrt f}.}
 \label{eq:lambda-threshold}
\end{equation}

\rev{This criterion describes the initial radial acceleration of a released test particle in a fixed background. It is not a force between two black holes. A particle held at rest at an arbitrary radius is generally supported by an external force; the static energy $E_{\rm stat}=\mu\sqrt f+qQ/r$ is therefore a potential for constrained static configurations, not a claim that a freely moving particle remains static. The sign of $-\dd E_{\rm stat}/\dd r$ agrees with the release criterion in Eq.~(\ref{eq:lambda-threshold}).}

\rev{A constant gauge shift $A_t\mapsto A_t+C$ changes the numerical value assigned to $\varepsilon$ by $\lambda C$ but leaves $\varepsilon-\lambda A_t$, its radial derivative, and the acceleration threshold invariant. In a Kiselev static patch, spatial infinity is not part of the static domain. It is therefore more operational to choose a finite reference radius $r_{\rm ref}$ within the patch and set $A_t(r_{\rm ref})=0$ by a constant gauge shift. The derivative and all force conditions remain unchanged.}

\rev{The admissible domain for static observers is the open interval $r_+<r<r_c$ when a cosmological horizon is present, or $r>r_+$ when the exterior is unbounded. The horizons themselves are not static-observer worldlines. For simple, nondegenerate horizons delimiting a static patch, $f'(r_+)>0$ and $f'(r_c)<0$. At a merger one of these derivatives may vanish, and the simple-horizon sign argument must not be used.}

\subsection{String-cloud threshold}

For the string-cloud geometry, Eq.~(\ref{eq:lambda-threshold}) becomes
\begin{equation}
 \lmin^2(r)=1+\frac{M^2(1-z^2)}{Q^2f(r)},
 \qquad z=\frac{Q}{Q_{\rm ext}}.
 \label{eq:string-lambda}
\end{equation}
For $0<z<1$, this gives $\lmin(r)>1/z$ at every finite exterior radius, while
\begin{equation}
 \inf_{r>r_+}\lmin(r)=\lim_{r\to\infty}\lmin(r)=\frac1z.
 \label{eq:string-infimum}
\end{equation}
The value $1/z$ is an infimum, not a finite-radius minimum. Hence a particle with $\lambda>1/z$ has outward initial acceleration sufficiently far from the black hole, whereas $\lambda\leq1/z$ does not. At extremality, $z=1$ and $\lmin=1$ throughout the static exterior.

\subsection{PFDM threshold and global numerical minimum}

\rev{For the logarithmic profile, set $B(r)=m-rm'=M+\frac\alpha2[\ln(r/\alpha)-1]$. Direct algebra gives}
\begin{equation}
 \rev{\lmin^2-1=\frac{B(r)^2-Q^2(1-\alpha/r)}{Q^2f(r)}.}
 \label{eq:pfdm-lambda-identity}
\end{equation}
\rev{On the positive-charge black-hole branch, $r_e>\alpha$. The numerator on the exterior is positive for a subextremal solution; Appendix~\ref{app:testcharge} proves this by comparison with the extremal charge. Consequently $\lmin(r)>1$ at every finite exterior radius. This makes $\lambda>1$ necessary, but it is not sufficient at a specified radius.}

\rev{For a subextremal PFDM solution, $\lmin(r)\to+\infty$ as $r\to r_+^+$ and also as $r\to\infty$, where $\lmin(r)\sim\alpha[2Q]^{-1}\ln(r/\alpha)$. It therefore attains a finite global minimum}
\begin{equation}
 \rev{\lstar\equiv\min_{r>r_+}\lmin(r)>1.}
 \label{eq:lambda-star}
\end{equation}
\rev{For a fixed particle charge-to-mass ratio  $\lambda$, at least one radius with outward initial acceleration exists if and only if $\lambda>\lstar$. No general uniqueness claim for the minimizing radius or for the number of boundary radii is required here; the reported values are obtained by a grid-assisted numerical minimization and are not presented as an analytic uniqueness theorem.}

\rev{For $M=0.5$, $\alpha=0.2$, and $Q=0.6$, we find}
\begin{equation}
 \rev{r_+=0.902435,\qquad r_{\lstar}=2.035128,
 \qquad \lstar=1.251125.}
 \label{eq:pfdm-lstar-example}
\end{equation}
\rev{Thus a particle with $\lambda=1.1$ has no outward-acceleration region in this example even though $\lambda>1$. On the extremal geometry with the same $M$ and $\alpha$, the radial limit at the degenerate horizon is instead}
\begin{equation}
 \rev{\lim_{r\to r_e^+}\lmin(r)
 =\sqrt{\frac{r_e-\alpha/2}{r_e-\alpha}}.}
 \label{eq:pfdm-extremal-limit}
\end{equation}
\rev{Numerically, $r_e=0.729389$ and this limit is $1.090366$. This is a radial limit on the extremal geometry; it is not the global minimum of the subextremal family and is not the Reissner-Nordstr\"om limit $\alpha\to0$.}

\begin{figure*}[t]
 \includegraphics[width=0.93\textwidth]{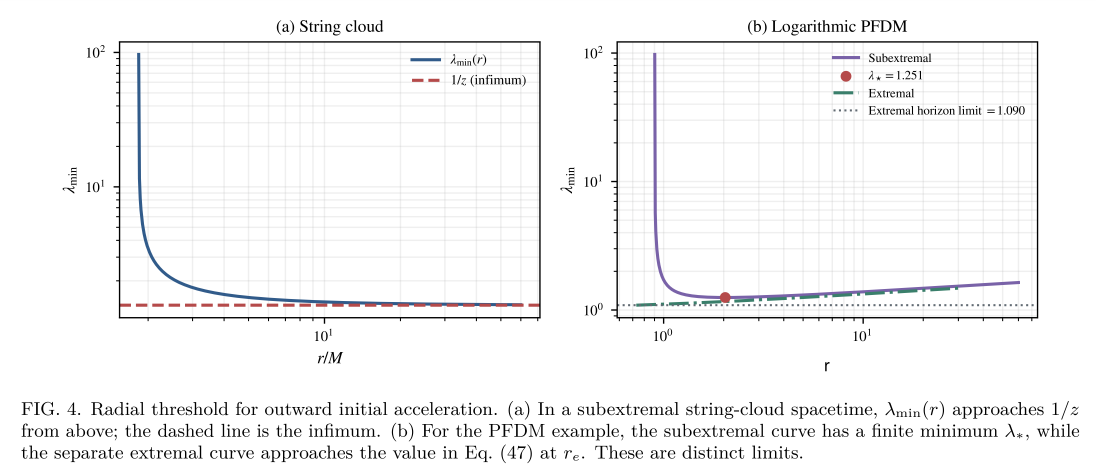}
 \caption{Radial threshold for outward initial acceleration. (a) In a subextremal string-cloud spacetime, $\lmin(r)$ approaches $1/z$ from above; the dashed line is the infimum. (b) For the PFDM example, the subextremal curve has a finite minimum $\lstar$, while the separate extremal curve approaches the value in Eq.~(\ref{eq:pfdm-extremal-limit}) at $r_e$. These are distinct limits.}
 \label{fig:thresholds}
\end{figure*}

\begin{figure}[t]
 \includegraphics[width=\columnwidth]{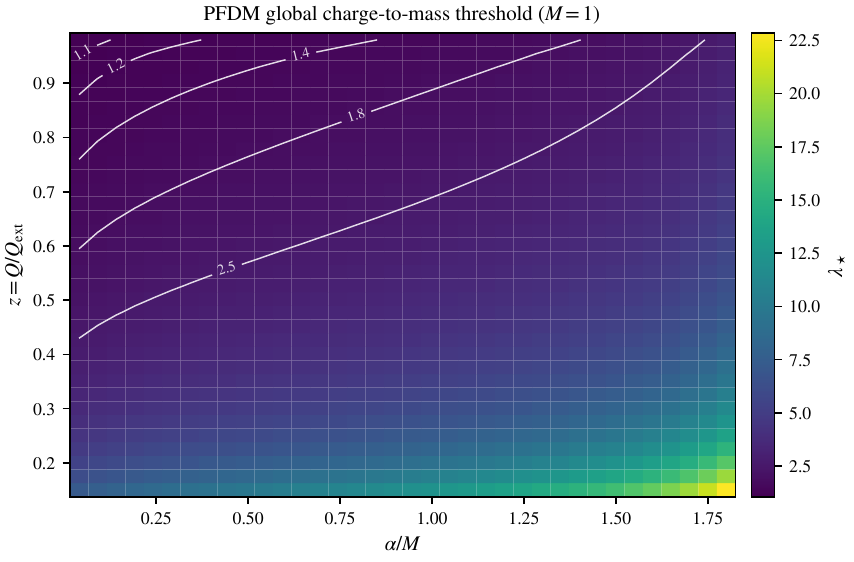}
 \caption{Numerically determined global threshold $\lstar$ for the PFDM profile at $M=1$, over the sampled plane $0.04\leq\alpha/M\leq1.8$ and $0.15\leq z=Q/Q_{\rm ext}\leq0.98$. Each cell is obtained from a one-dimensional minimization of $\lmin(r)$ outside the event horizon. The map is a numerical diagnostic over this domain, not an analytic proof of uniqueness of the minimizer.}
 \label{fig:lambda-map}
\end{figure}

\subsection{Neutral force-balance radius in the Kiselev static patch}

\rev{For a neutral body released from rest, Eq.~(\ref{eq:radial-acceleration}) gives $\ddot r=-f'/2=-N/r^2$. A force-balance radius satisfies $N(r_q)=0$. With $p=-3\wq$ and $A=1-b$, the function is}
\begin{equation}
 \ifhighlight\color{blue}\fi
 \begin{aligned}
 N(r)&=M-\frac{Q^2}{r}-\frac\alpha2(p-1)r^p,\\
 N'(r)&=\frac{Q^2}{r^2}
 -\frac\alpha2p(p-1)r^{p-1}.
 \end{aligned}
 \label{eq:kiselev-N}
\end{equation}
\normalcolor
\rev{The string-cloud parameter $b$ cancels explicitly from this equation. For $Q>0$, $N'$ has exactly one positive zero,}
\begin{equation}
 \rev{r_*^{(N)}=\left[\frac{2Q^2}{\alpha p(p-1)}\right]^{1/(p+1)},}
 \label{eq:kiselev-Nturn}
\end{equation}
\rev{so $N$ first increases and then decreases. For $Q=0$, $N$ is strictly decreasing. In a nondegenerate static patch bounded by simple horizons, $N(r_+)>0$ and $N(r_c)<0$. Thus exactly one zero $r_q$ lies in $(r_+,r_c)$, on the decreasing branch. At this point}
\begin{equation}
 \rev{f''(r_q)=\frac{2N'(r_q)}{r_q^2}<0,\qquad
 \left.\frac{\dd^2}{\dd r^2}(\mu\sqrt f)\right|_{r_q}<0.}
 \label{eq:kiselev-neutral-instability}
\end{equation}
\rev{
The neutral static-energy extremum is therefore a radial maximum and is unstable. Although $b$
does not enter $N(r)$ explicitly, it changes the horizon locations and the existence and
extent of the black-hole static patch. Whenever the assumed nondegenerate patch exists, the theorem above places exactly one neutral balance radius inside it. This uniqueness statement is restricted to
the Kiselev family and the nondegenerate static domain specified above.}
\begin{figure}[t]
 \includegraphics[width=\columnwidth]{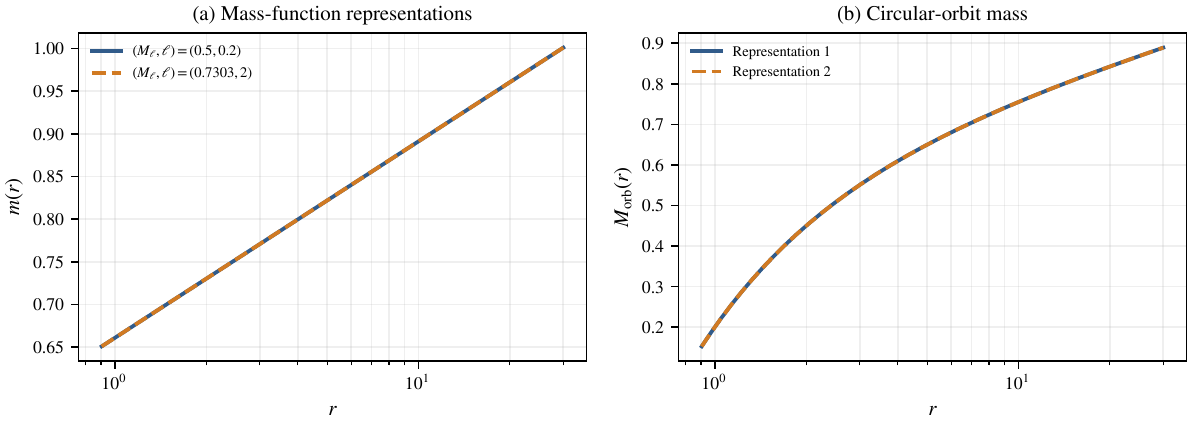}
 \caption{Reference-scale dependence of the logarithmic PFDM mass parameter. The two parameter pairs $(M_\ell,\ell)$ describe the same function $m(r)$ and the same circular-orbit mass function. This illustrates why the integration parameter alone cannot be identified with an observed orbital mass.}
 \label{fig:orbital-mass}
\end{figure}

\section{Conditional density normalizations}\label{sec:density}

\rev{The density relation in Eq.~(\ref{eq:density}) permits conditional normalizations of the metric parameters to representative environmental densities. Such a normalization matches the density at a selected reference radius only. It does not match the principal pressures or establish a common microscopic source. In particular, the Kiselev component is anisotropic except in the cosmological-constant limit, while the logarithmic PFDM profile has nonzero principal pressures despite its vanishing average pressure. The estimates below are therefore profile-based benchmarks, not fitted constraints on realistic halos or homogeneous quintessence.}

\paragraph{Kiselev component.}
In the limit $\wq\to-1$, the density becomes constant and $\alpha=\Lambda/3$, where
\begin{equation}
 \Lambda=\frac{3\Omega_\Lambda H_0^2}{c^2}
 \simeq1.09\times10^{-52}\,{\rm m}^{-2}
 \label{eq:lambda-cosmological}
\end{equation}
for the representative values $H_0=67.36\,{\rm km\,s^{-1}\,Mpc^{-1}}$ and $\Omega_\Lambda=0.6847$ [10]. \rev{This identification reproduces the cosmological-constant stress tensor only at $\wq=-1$. For other values, matching a density at one radius does not identify the static anisotropic Kiselev source with cosmological quintessence.}

\paragraph{PFDM logarithmic profile.}
For a reference density $\rho_{\rm DM}=0.4\,{\rm GeV\,cm^{-3}}$ at $r_0=8\,{\rm kpc}$, the adopted $r^{-3}$ profile gives
\begin{equation}
 \alpha_{\rm DM}=\frac{8\pi G\rho_{\rm DM}r_0^3}{c^2}
 \simeq2.0\times10^{14}\,{\rm m}.
 \label{eq:alphaDM}
\end{equation}
\rev{This prescription normalizes the density of the anisotropic logarithmic profile. It does not reproduce the stress tensor of a pressureless or nonrelativistic dark-matter halo, because the principal pressures in Eq.~(\ref{eq:pfdm-stress}) are not negligible relative to the density.}

\rev{The positive-charge horizon branch requires $\alpha<2M$. A direct comparison of $\alpha_{\rm DM}$ with an astronomical mass is meaningful only after specifying what mass combination the observation measures. For the logarithmic mass function the reference scale can be made explicit as}
\begin{equation}
 \rev{m(r)=M_\ell+\frac\alpha2\ln\!\left(\frac r\ell\right).}
 \label{eq:mass-reference-scale}
\end{equation}
\rev{Under a change of reference scale $\ell\mapsto\ell'$, the same geometry is represented by}
\begin{equation}
 \rev{M_{\ell'}=M_\ell+\frac\alpha2\ln\!\left(\frac{\ell'}\ell\right).}
 \label{eq:mass-scale-transform}
\end{equation}
\rev{Thus $M_\ell$ is convention-dependent. For a neutral test body, a circular timelike geodesic gives}
\begin{equation}
 \ifhighlight\color{blue}\fi
 \begin{aligned}
 \Omega^2&=\frac{f'}{2r},\qquad
 v_{\rm stat}^2=\frac{rf'}{2f},\\
 M_{\rm orb}(r)&\equiv r^3\Omega^2=\frac{r^2f'}2
 =m-rm'-\frac{Q^2}{r}.
 \end{aligned}
 \label{eq:orbital-mass}
\end{equation}
\normalcolor
\rev{These expressions describe a nonzero-angular-velocity timelike circular geodesic only where $f>0$, $f'>0$, and $2f-rf'>0$. They do not by themselves establish radial stability. Here $\Omega=d\phi/dt$ refers to the time coordinate in Eq.~(\ref{eq:metric}). Thus $M_{\rm orb}$ is defined with that time normalization and is not, without an observation model, a convention-independent measured mass.}
\rev{For the logarithmic profile this becomes}
\begin{equation}
 \rev{M_{\rm orb}(r)=M_\ell+
 \frac\alpha2\left[\ln\!\left(\frac r\ell\right)-1\right]
 -\frac{Q^2}{r}.}
 \label{eq:orbital-mass-pfdm}
\end{equation}
\rev{A measured orbital mass therefore constrains a radius-dependent combination, not $M_\ell$ alone. No orbit data or matching prescription sufficient to fit $(M_\ell,\alpha,\ell,Q)$ are supplied here. We consequently retain the astronomical comparison only as a nominal convention: setting $M_\alpha$ equal to the listed geometric mass and extending the profile to small radii places those parameter choices outside the positive-charge horizon branch. This algebraic statement is not a conclusion that the corresponding real black holes cannot possess an environment of this kind.}

\paragraph{String cloud.}
Requiring the string-cloud density not to exceed the same local reference density at $r_0$ gives
\begin{equation}
 b\lesssim\frac{8\pi G\rho_{\rm DM}r_0^2}{c^2}
 \simeq8\times10^{-7}.
 \label{eq:string-density-bound}
\end{equation}
The reference normalization is substantially below the illustrative values of $b$ used in the geometric figures. This comparison remains conditional on the $r^{-2}$ string-cloud profile and on the chosen reference radius.

\begin{figure}[t]
 \includegraphics[width=\columnwidth]{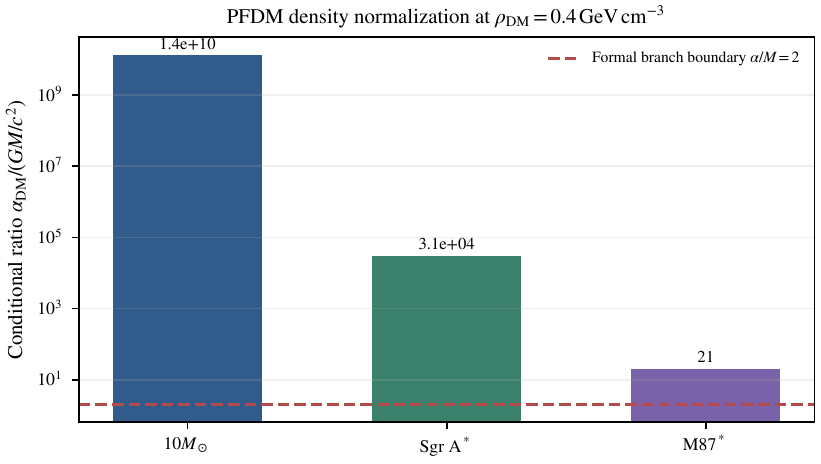}
 \caption{Nominal PFDM density-normalization ratios obtained by setting $M_\alpha$ equal to the listed geometric masses. The horizontal line marks the formal horizon threshold $\alpha/M=2$. The plot is an algebraic benchmark under this convention, not an orbit-based mass fit or an observational exclusion of the corresponding astrophysical environments.}
 \label{fig:density}
\end{figure}

\section{Cosmological consistency checks}\label{sec:cosmo}

\subsection{Astrophysical charge estimates}

In geometric units, the Reissner-Nordstr\"om extremal limit is $|Q|/M=1$. Charge accretion and discharge arguments suggest that astrophysical black holes are generally expected to lie far below this limit.
\rev{
For a negatively charged black hole, balancing Newtonian gravitational attraction against electrostatic repulsion of an electron gives the idealized electron-capture benchmark
\begin{equation}
 \frac{|Q|}{M}\lesssim
 \sqrt{4\pi\epsilon_0G}\,\frac{m_e}{e}
 \simeq4.9\times10^{-22},
 \qquad Q<0,
 \label{eq:electron-charge}
\end{equation}
where $Q/M$ is expressed in the geometric normalization, while the constants on the right-hand side are evaluated in SI units and $e>0$ is the elementary charge. This estimate assumes that the electron's initial kinetic energy is negligible. It is a sign-specific theoretical benchmark, not a population-wide observational limit or a general upper bound on the positive charge of a black hole in a plasma \cite{Zajacek2018}.} 
A Wald-type induced charge of order $10^{15}\,{\rm C}$ for Sgr~A$^*$ corresponds to $Q/M$ of order $10^{-12}$ [11]. A separate Galactic-centre analysis reports an observational upper limit $Q\lesssim3\times10^8\,{\rm C}$ [11]. These estimates refer to different charge-formation or inference assumptions and should not be conflated. They are also distinct from the test-particle threshold in Sec.~\ref{sec:forces}, which is a local condition in a prescribed background.

\subsection{Isotropy and homogeneous net charge}

\rev{An exactly homogeneous and isotropic electromagnetic field cannot contain a nonzero coherent spatial electric or magnetic vector: such a vector would select a preferred direction. Maxwell's Gauss law then excludes a nonzero homogeneous net charge density in this idealized setting. This symmetry argument does not rule out local inhomogeneous charge distributions, globally compensated configurations, or charge separation on sub-Hubble scales. The distinction is important because a local charged-black-hole solution is not itself a homogeneous charged FLRW source.}

\subsection{A homogeneous charged-dust toy model}

\rev{For clarity, we separate a fixed-mass dust model from a cosmologically varying black-hole mass model. Consider first a single-component Newtonian toy model in which each dust particle has fixed mass $m$ and charge $q$, and define the dimensionless charge parameter in SI units by}
\begin{equation}
 \rev{\lambda^2\equiv\frac{q^2}{4\pi\epsilon_0Gm^2}.}
 \label{eq:lambda-dust}
\end{equation}
\rev{A spherical shell of physical radius $R$ then obeys}
\begin{equation}
 \rev{\ddot R=-\frac{4\pi G}{3}\rho R(1-\lambda^2).}
 \label{eq:dust-acceleration}
\end{equation}
\rev{If particle number is conserved in a comoving volume, the comoving number density is constant and the physical number density scales as $a^{-3}$. With fixed $m$ and $q$, this gives $\rho\propto a^{-3}$ and constant $\lambda$. Integrating the shell equation yields the toy-model Friedmann relation}
\begin{equation}
 H^2=\frac{8\pi G}{3}\rho(1-\lambda^2)
 -\frac{Kc^2}{a^2}.
 \label{eq:charged-friedmann}
\end{equation}
\rev{The term proportional to $(1-\lambda^2)$ still scales as matter; it does not become a constant-density component. If $\lambda>1$, it contributes negatively to the effective matter term. The resulting curvature bookkeeping is a property of this single-fluid Newtonian toy model only. It is not a direct fit to Planck or BAO data and is not a constraint on a relativistic, charge-compensated cosmological model.}

\rev{For a separate variable-mass population with $m(a)\propto a^k$ and conserved comoving number, one instead has $\rho_{\rm BH}\propto a^{k-3}$ and}
\begin{equation}
 \rev{\dot\rho_{\rm BH}+3H\rho_{\rm BH}
 =kH\rho_{\rm BH}.}
 \label{eq:variable-mass-continuity}
\end{equation}
\rev{The right-hand side represents energy transfer into the black-hole population from another sector. A covariantly conserved total stress tensor requires a compensating term, for example $\dot\rho_X+3H(\rho_X+p_X)=-kH\rho_{\rm BH}$ for the sector $X$ that supplies the energy. If the black-hole charge is fixed while the mass changes, $q/m$ and hence $\lambda$ vary; holding $\lambda$ fixed instead requires a corresponding charge evolution. The fixed-mass charged-dust integration in Eq.~(\ref{eq:charged-friedmann}) cannot be combined with the variable-mass scaling without specifying this exchange sector and its dynamics.}

\begin{figure}[t]
 \includegraphics[width=\columnwidth]{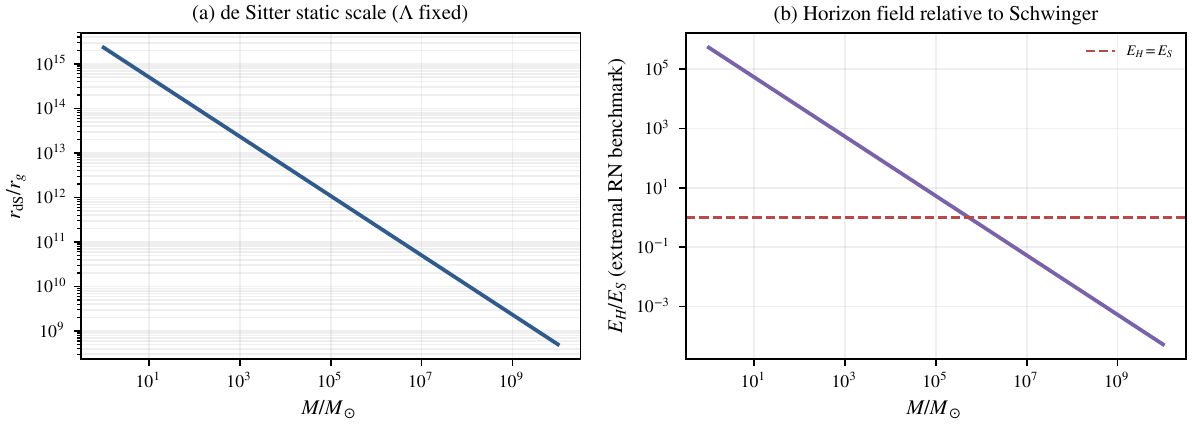}
 \caption{Two ancillary scales: (a) the de Sitter static radius $r_{\rm dS}=(3GM/\Lambda c^2)^{1/3}$ and (b) the extremal Reissner-Nordstr\"om horizon field relative to the Schwinger field. These are reference estimates and do not model charge production or discharge in the environmental geometries studied here.}
 \label{fig:cosmo-benchmarks}
\end{figure}

\section{Cosmologically coupled black holes and observational summaries}\label{sec:coupling}

\rev{The cosmological-coupling proposal concerns mass evolution with the scale factor and is independent of the static test-charge results. Published analyses have reported different constraints under different data selections, population assumptions, and mass-calibration systematics. For the combined elliptical-galaxy sample, Farrah \textit{et al.} report a 90\% interval $k=3.11^{+1.19}_{-1.33}$; a Gaussian approximation to that posterior has mean $3.09$ and standard deviation $0.76$ [1]. An analysis of nine JWST active galactic nuclei in red-sequence hosts reports $k=-0.16\pm0.39$ after its bias corrections [14]. Using a power-law-plus-peak population model
and a minimum formation mass of $2\Msun$, the analysis of Ref.~[15] gives $2\sigma$ upper limits $k<2.1$ for the standard log-uniform delay distribution and $k<1.1$ for the coupling-corrected distribution. The low-mass population in NGC~3201 has also been discussed in relation to a possible minimum mass set by the Tolman-Oppenheimer-Volkoff limit [16], while Gaia astrometric binaries provide an additional method-dependent constraint [17]. These results concern the mass-coupling hypothesis and do not test the WGC or the charged-test-particle condition derived above.}

Related theoretical work has examined cosmological horizon coupling and possible obstructions [18,19], dark-sector embeddings [20], and phenomenological fits to baryon acoustic oscillation distances [21,22]. Such constructions require a cosmological spacetime and matter model beyond the static ansatz in Eq.~(\ref{eq:metric}).

\rev{The parameter $\wq$ should not be identified with the present-day CPL parameter $w_0$. In the static geometry studied here, $\wq$ fixes the ratio of average principal pressure to density of an anisotropic source, $\bar p_q/\rho_q=\wq$. By contrast, CPL parameters describe a cosmological equation-of-state parametrization. No mapping between these descriptions is specified in this work. We therefore do not interpret numerical differences between $\wq$ and cosmological $w_0$ estimates as statistical pulls or as observational constraints on the Kiselev source.}

\begin{table*}[t]
\caption{Representative published summaries for the cosmological-coupling index $k$. The quoted confidence level and model assumptions are those of the corresponding analysis; the entries are not combined into a joint likelihood.}
\label{tab:kconstraints}
\begin{ruledtabular}
\small
\begin{tabular}{lcl}
Source & Data & Reported result \\
\colrule
Farrah et al. (2023) & Elliptical galaxies & $3.09\pm0.76$ (Gaussian approximation) \\
Lei et al. (2026) & 9 JWST AGN & $-0.16\pm0.39$ (bias-corrected) \\
Amendola et al. (2024) & LVK mergers & $k<2.1$ / $k<1.1$ ($2\sigma$; distinct assumptions) \\
Rodriguez (2023) & NGC 3201 & Model-dependent test of $k=3$ \\
Andrae-El-Badry (2023) & Gaia binaries & Method-dependent constraint \\
\end{tabular}
\end{ruledtabular}
\end{table*}

\section{Conclusions}\label{sec:conclusions}

\rev{We have examined horizon structure and test-particle dynamics in three prescribed charged, static geometries. Their environmental terms alter the extremal charge relative to the metric integration parameter $M$, so $Q/M>1$ can coexist with a horizon in the string-cloud and logarithmic PFDM branches. For the Kiselev-plus-string family, however, the horizon structure is two-sided: a black-hole static patch requires the critical-amplitude condition and the interval in Eq.~(\ref{eq:kiselev-horizon-criterion}). The ratio $z=Q/Q_C$ supplies only the upper merger condition. Degenerate merger surfaces, including the triple-root limit, are not nondegenerate static regions.}

\rev{The effective matter sources are fixed by the metric after the Maxwell contribution is separated. The Kiselev source is anisotropic for $-1<\wq<-1/3$, with $\wq=\bar p_q/\rho_q$; it becomes isotropic only in the cosmological-constant limit. The logarithmic PFDM source has $p_r=-\rho$ and $p_t=\rho/2$, so its vanishing average pressure does not make it pressureless. The names ``quintessence'' and ``perfect-fluid dark matter'' are therefore not used as identifications of the microscopic sources.}

\rev{
The charged-particle analysis characterizes the initial radial acceleration of a particle
released from rest relative to a static observer. For strings, the threshold of a subextremal
solution is strictly above $1/z$ at every finite exterior radius and approaches that value only asymptotically.

For every subextremal PFDM solution on the positive-charge black-hole branch, the
threshold exceeds unity throughout the finite exterior and attains a finite global minimum $\lstar>1$. Thus $\lambda>1$ is necessary, whereas $\lambda>\lstar$ is
necessary and sufficient for the existence of at least one radius with outward initial
acceleration. For the reference case $\alpha/M=0.4$ and $Q/M=1.2$, we obtain $\lstar\simeq1.251125$. On the extremal PFDM geometry with the same $M$ and $\alpha$, but with $Q=Q_{\rm ext}$, the radial horizon limit is $1.090366$. This limit is distinct from the global minimum of the reference
subextremal solution.

In the nondegenerate Kiselev static patch, a unique neutral balance radius exists and is radially unstable. None of these statements establishes a two-black-hole force law.
}

\rev{The density comparisons normalize selected profiles at reference radii. They do not fit the full stress tensor or determine the matter microphysics. For the logarithmic profile, a reference-scale transformation changes $M_\ell$ while leaving the geometry invariant, and circular motion measures the combination in Eq.~(\ref{eq:orbital-mass-pfdm}). Without orbital data and a matching prescription, the astronomical mass comparison is only conditional. Likewise, the homogeneous charged-dust calculation is a Newtonian toy model; it is not a relativistic cosmological fit or a direct observational constraint.}

A two-centre solution, a covariant matter action, a specified gauge normalization, and a charged spectrum are needed before geometric extremality bounds can be connected to a WGC statement. A realistic environmental interpretation of the logarithmic profile would additionally require an inner matching prescription, an outer cutoff or transition, and a check of the corresponding junction stresses. Rotating solutions, time-dependent embeddings, and a self-consistent perturbation analysis are further extensions rather than consequences of the present static calculation.

\appendix

\section{Stress tensor and conservation identities}\label{app:stress}

For the metric in Eq.~(\ref{eq:metric}), after subtracting the Maxwell contribution, the environmental Einstein equations can be written as
\begin{align}
 8\pi\rho&=\frac{1-f-rf'}{r^2}-\frac{Q^2}{r^4},\label{eq:app-rho}\\
 8\pi p_r&=\frac{rf'+f-1}{r^2}+\frac{Q^2}{r^4},\label{eq:app-pr}\\
 8\pi p_t&=\frac{f''}{2}+\frac{f'}r-\frac{Q^2}{r^4}.
 \label{eq:app-pt}
\end{align}
Substitution of $f=1-2m/r+Q^2/r^2$ gives Eqs.~(\ref{eq:density}) and (\ref{eq:environment-pressures}). For each component, Eq.~(\ref{eq:environment-conservation}) reduces to
\begin{equation}
 p_r'+\frac2r(p_r-p_t)=0,
\end{equation}
which is satisfied identically.

\rev{Within this one-function metric ansatz, an isotropic environmental source would require $p_t=p_r$, hence $m''=2m'/r$. Its solution is $m=M_0+Cr^3$, giving $\rho=3C/(4\pi)$ and $p_r=p_t=-\rho$. Thus the only nonvacuum isotropic source admitted by this restricted ansatz is of cosmological-constant type. This statement does not apply to general static spherical metrics with independent temporal and radial metric functions.}

\section{The radial diagnostic and quasi-local mass}\label{app:diagnostic}

At an outer horizon, Eq.~(\ref{eq:proxy-horizon-identity}) yields $|m(r_+)|\geq Q$. On the string-cloud and Kiselev-plus-string branches with $M>0$, $b\geq0$, and $\alpha>0$, $m(r_+)>0$. On the PFDM branch, $r_e>\alpha$ and Eq.~(\ref{eq:pfdm-extremal}) gives $m(r_e)=r_e-\alpha/2>0$; since $m'\geq0$, the same sign holds at the outer horizon and throughout the exterior. Thus $m(r_+)\geq Q$ and $m(r)\geq Q$ for $r\geq r_+$, establishing Eq.~(\ref{eq:proxy-bound}). The argument is a bound on the constructed radial diagnostic, not a statement about a force between compact objects.

The Misner-Sharp mass follows directly from $M_{\rm MS}=r(1-f)/2$ and is given by Eq.~(\ref{eq:ms-mass}). The combination $N=r^2f'/2$ instead controls the sign of the radial acceleration of a neutral particle instantaneously at rest. These three quantities answer different geometric or dynamical questions.

\section{Kiselev horizon branches and triple-root boundary}\label{app:kiselev}

Let $A=1-b$ and $p=-3\wq\in(1,3)$. The stationary equation is $H(r)=M-Ar+\alpha(p+1)r^p/2=0$. For $A>0$, its derivative
\begin{equation}
 \begin{aligned}
 H'(r)&=-A+\frac{\alpha p(p+1)}2r^{p-1},\\
 H''(r)&=\frac{\alpha p(p+1)(p-1)}2r^{p-2}>0.
 \end{aligned}
\end{equation}
has one zero, so $H$ has a unique minimum. Solving $H=H'=0$ gives Eq.~(\ref{eq:acrit}). If $0<\alpha<\alpha_{\rm crit}$, the two zeros $r_a<r_b$ divide the positive axis into intervals where $g$ increases, decreases, and increases. Since $g(0^+)=0$ and $g(\infty)=+\infty$, a positive level $Q^2$ intersects all three intervals if and only if $\max\{0,g(r_b)\}<Q^2<g(r_a)$. This proves Eq.~(\ref{eq:kiselev-horizon-criterion}).

At a double root $r_i$, $f''(r_i)=-g''(r_i)/r_i^2$. Thus $r_a$ is the inner-event merger and $r_b$ is the event-cosmological merger when $g(r_b)>0$. At $\alpha=\alpha_{\rm crit}$ the stationary radii coalesce at
\begin{equation}
 r_u=\frac{pM}{A(p-1)},\qquad
 Q_u^2=g(r_u)=\frac{p^2M^2}{A(p^2-1)}.
\end{equation}
\rev{If the charge also satisfies $Q^2=Q_u^2$, the horizon at $r=r_u$ is a triple root, with $f=f'=f''=0$ and $f'''\neq0$. For $\alpha=\alpha_{\rm crit}$ but $Q^2\neq Q_u^2$, the unique positive horizon is simple.}

\section{Charged test-particle dynamics and PFDM inequalities}\label{app:testcharge}

\rev{The charged-particle dynamics follows from the action
\begin{equation}
 S=-\mu\int ds-q\int A_\mu dx^\mu,
 \qquad A_t=\frac{Q}{r}.
\end{equation}
With the metric signature $(+--)$, the canonical
momentum conjugate to $t$ is
\begin{equation}
 P_t=-\mu f\dot t-qA_t.
\end{equation}
With the sign chosen to reproduce
$E=\mu f\dot t$ in the neutral limit,
the conserved canonical energy is
$E=-P_t$, so that
\begin{equation}
 \varepsilon\equiv\frac{E}{\mu}
 =f\dot t+\lambda\frac{Q}{r},
 \qquad \lambda=\frac{q}{\mu},
\end{equation}
in agreement with Eq.~(\ref{eq:canonical-energy}).
The normalization $u^\mu u_\mu=1$ gives
Eq.~(\ref{eq:radial-geodesic}).
Differentiating and imposing instantaneous rest gives
Eq.~(\ref{eq:radial-acceleration}) and the threshold
in Eq.~(\ref{eq:lambda-threshold}).}

For PFDM, write $B=m-rm'=M+\frac\alpha2[\ln(r/\alpha)-1]$ and $h(r;Q)=B^2-Q^2(1-\alpha/r)$. On the exterior of the extremal branch, $r\geq r_e>\alpha$, so $1-\alpha/r>0$ and $h$ is minimized, at fixed $r$, by the largest allowed charge, $Q=Q_{\rm ext}$. Define
\begin{equation}
 \phi(r)=r h(r;Q_{\rm ext})=rB^2-Q_{\rm ext}^2(r-\alpha).
\end{equation}
At $r=r_e$, extremality gives $B(r_e)=r_e-\alpha$ and $Q_{\rm ext}^2=r_e(r_e-\alpha)$, so $\phi(r_e)=\phi'(r_e)=0$. Moreover,
\begin{equation}
 \phi''(r)=B'(2B+\alpha)
 =\frac{\alpha}{2r}(2B+\alpha)>0\qquad (r\geq r_e).
\end{equation}
It follows that $h(r;Q_{\rm ext})>0$ for $r>r_e$, and therefore $h(r;Q)>0$ for subextremal $Q<Q_{\rm ext}$. Equation~(\ref{eq:pfdm-lambda-identity}) then gives $\lmin^2>1$.

To fix the sign, note that
\begin{equation}
 N'(r)=\frac{\alpha}{2r}+\frac{Q^2}{r^2}>0.
\end{equation}
At a simple subextremal event horizon, $N(r_+)=r_+^2f'(r_+)/2>0$. Hence $N(r)>0$ for $r>r_+$, and $\lmin=N/(Q\sqrt f)>0$. Consequently, $\lmin>1$ throughout the finite exterior.

For the extremal horizon limit, expand $f$ and $N$ around $r_e$. Since $f(r_e)=f'(r_e)=0$ and $f''(r_e)>0$, one obtains
\begin{equation}
 \lim_{r\to r_e^+}\lmin(r)^2
 =\frac{r_e-\alpha/2}{r_e-\alpha},
\end{equation}
which is Eq.~(\ref{eq:pfdm-extremal-limit}). This radial limit is not the minimum of the subextremal family.

\section{Numerical methods}\label{app:numerics}

The samples are numerical realizations of the analytically admissible branches, rather than an independent proof of the branch classification over the full parameter space. The principal numerical settings are summarized in Table~\ref{tab:numerics}.

\begin{table*}[t]
\caption{Numerical settings used for the accompanying seeded checks.}
\label{tab:numerics}
\begin{ruledtabular}
\small
\begin{tabular}{lc}
Quantity & Setting \\
\colrule
Random seed & $20261109$ \\
String samples & 1000 / 1000 accepted \\
PFDM samples & 1000 / 1000 accepted \\
Kiselev + strings & 1000 accepted; $\leq5000$ proposals \\
String $b$ & $U(0,0.8)$ \\
String/PFDM $z$ & $U(0.05,0.995)$ \\
PFDM $\alpha/M$ & $U(0.01,1.9)$ \\
Kiselev $p$ & $U(1.05,2.95)$ \\
Kiselev $b$ & $U(0,0.8)$ \\
Kiselev $\alpha/\alpha_{\rm crit}$ & $U(0.03,0.95)$ \\
Kiselev $Q^2$ & $L+u(U-L)$, $u\sim U(0.03,0.97)$ \\
Root bracketing & Brent, split at analytic stationary radii \\
\end{tabular}
\end{ruledtabular}
\end{table*}
\rev{
For strings and PFDM, the charge is parameterized by the sampled subextremal ratio $z$. For the
Kiselev family, the stationary radii are obtained first. Define
\begin{equation}
 L=\max(0,G_N),\qquad U=Q_C^2.
\end{equation}
We draw $u\sim U(0.03,0.97)$ and set
\begin{equation}
 Q^2=L+u(U-L).
\end{equation}
The sampling therefore excludes the nearest three percent of the squared-charge interval at each end. When $L=0$, the lower endpoint is zero charge, not necessarily a horizon-merger boundary. Root searches are performed on the separate monotonic radial intervals. Degenerate roots are tested separately through the stationary-point conditions. The samples are not an astrophysical population and are not sampled uniformly over the full squared-charge interval.}

\rev{The maximum absolute horizon residuals over the 1000 accepted samples are $3.30\times10^{-11}$ (string cloud), $2.98\times10^{-14}$ (PFDM), and $4.64\times10^{-11}$ (Kiselev plus strings). The maximum relative residual in the Kiselev stationary equation, scaled by the largest term in that equation, is $3.44\times10^{-14}$. In a small subset with $p$ close to unity, the outer stationary radius becomes very large and the unscaled absolute residual is correspondingly less informative; the relative residual is the appropriate diagnostic there.}

\rev{The PFDM global threshold is computed by minimizing $\lmin(r)$ in a logarithmic coordinate measured from the outer horizon, with a coarse scan used to bracket the lowest candidate and a bounded scalar minimizer used for refinement. The output map covers $0.04\leq\alpha/M\leq1.8$ and $0.15\leq z\leq0.98$.}

The reference sample for the PFDM threshold is $(M,\alpha,Q)=(0.5,0.2,0.6)$. The scale-invariance check uses $(M_{\ell_1},\ell_1)=(0.5,0.2)$ and $(M_{\ell_2},\ell_2)=(0.7302585,2.0)$; the maximum absolute differences in $m(r)$ and $M_{\rm orb}(r)$ on the plotted interval are below $3\times10^{-16}$ in the stored floating-point calculation.

The accompanying symbolic and regression tests check the general double-horizon charge identity, the canonical energy derived from the chosen particle action, and the first-order Kiselev expansion. The latter is also tested at masses different from unity. These checks verify mathematical identities and numerical implementations; they do not automatically parse or certify all equations

\section{Small-parameter limits and reference values}\label{app:limits}

For $b\ll1$, Eq.~(\ref{eq:string-extremal}) yields Eq.~(\ref{eq:string-expansion}). For the Kiselev component with $b=0$, an expansion around the Reissner-Nordstr\"om limit gives
\rev{\begin{equation}
 \frac{Q_C}{M}-1
 \simeq \frac{\alpha M^{p-1}}{2},
 \qquad p=-3\wq.
\end{equation}}
\rev{The dimensionless Kiselev amplitude is $\alpha M^{p-1}$. For $p=2$, it is $\alpha M$.} For PFDM the leading logarithmic term is Eq.~(\ref{eq:pfdm-expansion}). Table~\ref{tab:expansion} compares the non-expanded numerical values with the corresponding leading approximations. Here ``non-expanded'' means that the defining extremality equations are solved numerically; the displayed decimals are rounded values, not exact symbolic numbers.

\begin{table*}[t]
\caption{Numerical extremality shifts at $M=1$.
For strings the expansion parameter is $b$;
for PFDM it is $\alpha/M$;
for the $p=2$ Kiselev case it is
$\alpha M^{p-1}=\alpha M$.
The non-expanded numerical entries are obtained
by solving the defining extremality equations;
the displayed decimals are rounded values,
not exact symbolic numbers.}
\label{tab:expansion}
\begin{ruledtabular}
\small
\begin{tabular}{ccccccc}
Parameter & Strings (num.) & Strings (series) & Kiselev (num.) & Kiselev (series) & PFDM (num.) & PFDM (series) \\
\colrule
$10^{-4}$ & $5.0004\times10^{-5}$ & $5.0004\times10^{-5}$ & $5.0010\times10^{-5}$ & $5.0000\times10^{-5}$ & $4.6054\times10^{-4}$ & $4.6052\times10^{-4}$ \\
$10^{-3}$ & $5.0038\times10^{-4}$ & $5.0038\times10^{-4}$ & $5.0100\times10^{-4}$ & $5.0000\times10^{-4}$ & $3.4557\times10^{-3}$ & $3.4539\times10^{-3}$ \\
$10^{-2}$ & $5.0378\times10^{-3}$ & $5.0375\times10^{-3}$ & $5.1030\times10^{-3}$ & $5.0000\times10^{-3}$ & $2.3153\times10^{-2}$ & $2.3026\times10^{-2}$ \\
\end{tabular}
\end{ruledtabular}
\end{table*}

\section{Unit conversions and reference inputs}\label{app:units}

The conversions used in the conditional density estimates are $GM_\odot/c^2=1476.7\,{\rm m}$, $\Lambda=1.089\times10^{-52}\,{\rm m}^{-2}$, and $\rho_{\rm DM}=0.4\,{\rm GeV\,cm^{-3}}\simeq7.13\times10^{-22}\,{\rm kg\,m^{-3}}$. At $r_0=8\,{\rm kpc}$ this gives $\alpha_{\rm DM}\simeq2.0\times10^{14}\,{\rm m}$ and $b_{\max}\simeq8.1\times10^{-7}$. The representative masses used for the conditional display are $10\Msun$, $4.3\times10^6\Msun$ for Sgr~A$^*$, and $6.5\times10^9\Msun$ for M87$^*$. The horizon condition $\alpha<2M$ corresponds nominally to $M\gtrsim6.8\times10^{10}\Msun$ for this normalization. This last comparison assumes $M_\alpha$ equals the quoted geometric mass and is not an orbit-based determination.

\section{Schwinger-field benchmark}\label{app:schwinger}

\rev{For a Reissner-Nordstr\"om reference geometry, let $M_{\rm SI}$ denote the mass in kilograms and $Q_{\rm SI}$ the charge in coulombs.
Define the dimensionless charge fraction
$\epsilon$, with $|\epsilon|\leq1$, by
\begin{equation}
 Q_{\rm SI}
 =\epsilon\sqrt{4\pi\epsilon_0G}\,M_{\rm SI}.
\end{equation}
The outer horizon radius is
\begin{equation}
 r_+=\frac{GM_{\rm SI}}{c^2}
 \left(1+\sqrt{1-\epsilon^2}\right).
\end{equation}
The magnitude of the horizon electric field is
\begin{equation}
 E_H=\frac{|Q_{\rm SI}|}
 {4\pi\epsilon_0r_+^2}.
\end{equation}
At extremality, $|\epsilon|=1$, and for
$M_{\rm SI}=\Msun$, this gives
$E_H\simeq7.06\times10^{23}\,
{\rm V\,m^{-1}}$, so that
\begin{equation}
 \frac{E_H}{E_S}
 \simeq5.3\times10^5
 \left(\frac{\Msun}{M_{\rm SI}}\right),
\end{equation}
where $E_S$ is the Schwinger field. This is only a reference scale; no discharge rate or quantum-field calculation in the three environmental metrics is performed here.}

\section{Homogeneous isotropy and charge}

A spatial electric or magnetic vector is not invariant under all rotations. Therefore an exactly homogeneous and isotropic electromagnetic field has no coherent spatial vector component. Gauss's law then forbids a nonzero homogeneous net charge density in this idealized state. This symmetry statement does not preclude local inhomogeneous charges or globally compensated configurations.

\section{A regularized logarithmic profile as future work}\label{app:regularized}

\rev{A possible starting point for separating the central mass from an environmental contribution is a regularized profile in a finite radial domain,}
\begin{equation}
 \rev{m(r)=M_{\rm BH}+\frac\alpha2
 \ln\!\left(\frac{r+r_s}{r_{\rm in}+r_s}\right).}
 \label{eq:regularized-profile}
\end{equation}
\rev{In the region where this expression is used, it gives}
\begin{equation}
 \rev{\rho(r)=\frac{\alpha}{8\pi r^2(r+r_s)}.}
 \label{eq:regularized-density}
\end{equation}
\rev{This is a proposed profile for a future matching study, not a completed halo model. An inner vacuum region, an outer transition or cutoff, continuity of the metric, and any surface-layer stresses must be specified before the profile can be assigned a realistic astrophysical interpretation. The horizon and test-particle analysis should then be repeated for the matched geometry.}

\section*{Acknowledgments}
AI-assisted writing statement: The authors employed AI-based utilities during the drafting phase solely to polish the language, rectify grammatical errors, and facilitate \LaTeX\ typesetting. Subsequent to utilizing these resources, the authors carefully reviewed all sections, made necessary modifications, and bear sole responsibility for the content of the submitted article. It is hereby declared that every analytical outcome, conceptual framework, and interpretative assertion originates from the author's independent scholarly work.


\begin{thebibliography}{99}
\bibitem{Farrah2023} D. Farrah \textit{et al.}, Observational evidence for cosmological coupling of black holes and its implications for an astrophysical source of dark energy, Astrophys. J. Lett. \textbf{944}, L31 (2023), doi:10.3847/2041-8213/acb704, arXiv:2302.07878.
\bibitem{WGC2007} N. Arkani-Hamed, L. Motl, A. Nicolis, and C. Vafa, The string landscape, black holes and gravity as the weakest force, J. High Energy Phys. \textbf{06}, 060 (2007), doi:10.1088/1126-6708/2007/06/060, arXiv:hep-th/0601001.
\bibitem{WGCReview2023} D. Harlow, B. Heidenreich, M. Reece, and T. Rudelius, The weak gravity conjecture: A review, Rev. Mod. Phys. \textbf{95}, 035003 (2023), doi:10.1103/RevModPhys.95.035003, arXiv:2201.08380.
\bibitem{Sadeghi2023} J. Sadeghi, S. Noori Gashti, M. R. Alipour, and M. A. S. Afshar, Can black holes cause cosmic expansion?, arXiv:2305.12545 (2023).
\bibitem{Alipour2023a} M. R. Alipour, J. Sadeghi, and M. Shokri, WGC and WCCC of black holes with quintessence and cloud strings in RPS space, Nucl. Phys. B \textbf{990}, 116184 (2023).
\bibitem{Alipour2023b} M. R. Alipour, J. Sadeghi, and M. Shokri, WGC and WCC for charged black holes with quintessence and cloud of strings, Eur. Phys. J. C \textbf{83}, 640 (2023), doi:10.1140/epjc/s10052-023-11811-5.
\bibitem{Shaymatov2021} S. Shaymatov, B. Ahmedov, and M. Jamil, Testing the weak cosmic censorship conjecture for a Reissner-Nordstr\"om-de Sitter black hole surrounded by perfect fluid dark matter, Eur. Phys. J. C \textbf{81}, 588 (2021), doi:10.1140/epjc/s10052-021-09398-w.
\bibitem{Sadeghi2024} J. Sadeghi and S. Noori Gashti, Reissner-Nordstr\"om black holes surrounded by perfect fluid dark matter: testing the viability of weak gravity conjecture and weak cosmic censorship conjecture simultaneously, Phys. Lett. B \textbf{853}, 138651 (2024), doi:10.1016/j.physletb.2024.138651, arXiv:2404.15998.
\bibitem{Atamurotov2022} F. Atamurotov, U. Papnoi, and K. Jusufi, Shadow and deflection angle of charged rotating black hole surrounded by perfect fluid dark matter, Class. Quantum Grav. \textbf{39}, 025014 (2022), doi:10.1088/1361-6382/ac3e76, arXiv:2104.14898.
\bibitem{Planck2020} Planck Collaboration, N. Aghanim \textit{et al.}, Planck 2018 results. VI. Cosmological parameters, Astron. Astrophys. \textbf{641}, A6 (2020), doi:10.1051/0004-6361/201833910.
\bibitem{Zajacek2018} M. Zaja\v cek \textit{et al.}, On the charge of the Galactic centre black hole, Mon. Not. R. Astron. Soc. \textbf{480}, 4408 (2018), doi:10.1093/mnras/sty2182, arXiv:1808.07327.
\bibitem{Baushev2009} A. N. Baushev and P. Chardonnet, Electric charge estimation of a new-born black hole, arXiv:0905.4071 (2009).
\bibitem{Gibbons1975} G. W. Gibbons, Vacuum polarization and the spontaneous loss of charge by black holes, Commun. Math. Phys. \textbf{44}, 245 (1975), doi:10.1007/BF01609829.
\bibitem{Lei2026} L. Lei \textit{et al.}, Black holes in red-sequence galaxies at $z\sim0.7$-$2.5$: not a source of dark energy, but remnants of little red dots, Astron. Astrophys. \textbf{714}, L2 (2026), doi:10.1051/0004-6361/202661516.
\bibitem{Amendola2024} L. Amendola, D. C. Rodrigues, S. Kumar, and M. Quartin, Constraints on cosmologically coupled black holes from gravitational wave observations and minimal formation mass, Mon. Not. R. Astron. Soc. \textbf{528}, 2377 (2024), doi:10.1093/mnras/stae143.
\bibitem{Rodriguez2023} C. L. Rodriguez, Constraints on the cosmological coupling of black holes from the globular cluster NGC 3201, Astrophys. J. Lett. \textbf{947}, L12 (2023), doi:10.3847/2041-8213/acc9b6, arXiv:2302.12386.
\bibitem{Andrae2023} R. Andrae and K. El-Badry, Constraints on the cosmological coupling of black holes from Gaia, Astron. Astrophys. \textbf{673}, L10 (2023), doi:10.1051/0004-6361/202346350, arXiv:2305.01307.
\bibitem{Faraoni2024} V. Faraoni and M. Rinaldi, Black hole event horizons are cosmologically coupled, Phys. Rev. D \textbf{110}, 063553 (2024), doi:10.1103/PhysRevD.110.063553, arXiv:2407.14549.
\bibitem{Avelino2023} P. P. Avelino, Can gravitational vacuum condensate stars be a dark energy source?, J. Cosmol. Astropart. Phys. \textbf{08}, 005 (2023), doi:10.1088/1475-7516/2023/08/005, arXiv:2303.06630.
\bibitem{Wu2026} C.-H. Wu, Y. Chu, and Y.-P. Hu, Cosmological coupled black holes immersed in dark sector, Eur. Phys. J. C \textbf{86}, 288 (2026), doi:10.1140/epjc/s10052-026-15533-2, arXiv:2603.24609.
\bibitem{Croker2024} K. S. Croker \textit{et al.}, DESI dark energy time evolution is recovered by cosmologically coupled black holes, arXiv:2405.12282 (2024).
\bibitem{DESI2025} DESI Collaboration, DESI DR2 results. II. Measurements of baryon acoustic oscillations and cosmological constraints, Phys. Rev. D \textbf{112}, 083515 (2025), doi:10.1103/tr6y-kpc6, arXiv:2503.14738.
\bibitem{Kiselev2003} V. V. Kiselev, Quintessence and black holes, Class. Quantum Grav. \textbf{20}, 1187 (2003), doi:10.1088/0264-9381/20/6/310, arXiv:gr-qc/0210040.
\bibitem{Visser2020} M. Visser, The Kiselev black hole is neither perfect fluid, nor is it quintessence, Class. Quantum Grav. \textbf{37}, 045001 (2020), doi:10.1088/1361-6382/ab60b8, arXiv:1908.11058.
\bibitem{DariescuLungu2026} M.-A. Dariescu and V. Lungu, Epicyclic motion of charged particles around a weakly magnetized Kiselev black hole, Eur. Phys. J. C \textbf{86}, 296 (2026), doi:10.1140/epjc/s10052-026-15550-1, arXiv:2601.07736.
\bibitem{SilvaEtAl2026} M. V. de S. Silva, T. M. Crispim, R. R. Landim, G. J. Olmo, and D. S\'aez-Chill\'on G\'omez, Tidal forces around the Letelier-Alencar cloud of strings black hole, Phys. Rev. D \textbf{113}, 064052 (2026), doi:10.1103/3pj3-mx38, arXiv:2511.21604.
\bibitem{Ali2026} R. Ali, Tunneling properties and emission energy from a cloud of strings around charged black holes with bumblebee gravity, Ann. Phys. (Berlin) \textbf{538}, e70313 (2026), doi:10.1002/andp.70313.
\bibitem{BarbosaSantos2026} L. G. Barbosa and L. C. N. Santos, Black string immersed in perfect fluid dark matter, Phys. Lett. B \textbf{879}, 140620 (2026), doi:10.1016/j.physletb.2026.140620.
\bibitem{OliveiraSchmidt2026} M. D. de Oliveira and A. G. M. Schmidt, Exact solution of the Klein-Gordon equation in a Kiselev black hole background, Eur. Phys. J. C \textbf{86}, 1097 (2026), doi:10.1140/epjc/s10052-026-16325-4, arXiv:2609.04467.
\end{thebibliography}
\end{document}